\documentclass[cits]{PoS}
\usepackage[authoryear,square]{natbib}
\bibpunct{(}{)}{;}{a}{}{,}

\title{Galactic and Magellanic Evolution with the SKA}

\ShortTitle{Galactic and Magellanic Evolution}

\author{\speaker{Naomi M.\ McClure-Griffiths}$^1$,
Sne\v{z}ana Stanimirovi\'{c}$^2$, 
Claire E.\ Murray$^2$,
Di Li$^3$,
John M.\ Dickey$^4$,
Enrique V\'{a}zquez-Semadeni$^5$,
Josh E.\ G.\ Peek$^6$, 
Mary Putman$^6$, 
Susan E.\ Clark$^6$,
Marc-Antoine Miville-Desch\^{e}nes$^7$,
Joss Bland-Hawthorn$^8$,
Lister Staveley-Smith$^9$,
on behalf of the H~{\sc i} Science Working Group
\\
$^1$CSIRO Astronomy \& Space Science, Australia; $^2$University of
Wisconsin-Madison, USA; $^3$National Astronomical Observatories of
China, China; $^4$University of Tasmania, Australia; $^5$Centro de
Radioastronom\'{i}a y Astrof\'{i}sica, UNAM, Morelia, Mexico;
$^6$Columbia University, USA; $^7$CNRS - Institut d'Astrophysique
Spatiale, Universit\'{e} Paris-XI, Orsay, France; $^8$Sydney Institute

for Astrophysics, University of Sydney, Australia; $^9$International
Centre for Radio Astronomy Research , University of Western Australia,
Australia
\\
 E-mail: \email{naomi.mcclure-griffiths@csiro.au}}

\abstract{ 
  As we strive to understand how galaxies evolve 
it is crucial that we resolve physical processes
and test emerging theories
in nearby systems that we can observe in great detail. Our own Galaxy, the Milky
  Way, and the nearby Magellanic Clouds provide unique windows into
  the evolution of galaxies, each with its own metallicity and star
  formation rate.  These laboratories allow us to study with more
  detail than anywhere else in the Universe how galaxies acquire fresh
  gas to fuel their continuing star formation, 
how they exchange gas with the surrounding intergalactic medium, 
and turn warm, diffuse gas into molecular clouds and
  ultimately stars.  The $\lambda$21-cm line of atomic hydrogen
  (H~{\sc i}) is an excellent tracer of these physical processes.
  With the SKA we will finally have the combination of surface
  brightness sensitivity, point source sensitivity and angular
  resolution to transform our understanding of the
  evolution of gas in the Milky Way, all the way from the halo down to
  the formation of individual molecular clouds.}

\FullConference{
Advancing Astrophysics with the Square Kilometre Array\\
June 8-13, 2014\\
Giardini Naxos, Italy}

\newcommand{\skipthis}[1]{}

\newcommand\apj{ApJ}
\newcommand\apjs{ApJS}

\newcommand\araa{ARA\&A}
\newcommand\aj{AJ}
\newcommand\apjl{ApJL}
\newcommand\mnras{MNRAS}
\newcommand\nat{Nature}
\newcommand\aap{A\&A}
\newcommand\HI{H~{\sc i}}

\newcommand{\kms}{${\rm km~s^{-1}}$}

\newcommand\arcdeg{\mbox{$^\circ$}}%
\newcommand\arcmin{\mbox{$^\prime$}}%
\newcommand\arcsec{\mbox{$^{\prime\prime}$}}%

\begin{document}


\section{Introduction}

The next decades offer us the opportunity to revolutionize our
understanding of how galaxies form and evolve.  The current paradigm is that
galaxies form at the nexus between colliding streams of cold dark
matter, where new stellar systems form through bursts of star
formation \citep[SF; e.g.][]{vogelsberger14}. Strong winds driven by SF can force gas out of galaxies \citep{veilleux05} and
conversely galaxies must accrete new material to sustain their SF rates.  In addition, there is mounting evidence from
numerical simulations that molecular clouds form at the interfaces of
colliding streams of warm atomic gas,
at least under conditions similar to those in the present Milky Way.
Thus, there appears to exist a continuous gas
flow from extragalactic scales down to stellar scales.  

The Milky Way (MW) and Magellanic System are ideal laboratories for
studying the evolution of gas in galaxies. These systems have a wide
range of physical conditions, including high and low interstellar
heavy element abundance and SF rates.  They are close, so
close that the SKA will achieve spatial resolution finer than one
parsec, in some cases as small as an AU. With this extreme spatial
resolution, in combination with high spectral resolution, we can
resolve many of the open questions about galaxy evolution: how
galaxies acquire fresh gas, how they feed gas to local environments
and how they turn warm, diffuse gas into cold, dense gas and
ultimately form stars.  
In addition, the Magellanic Stream (MS), Large Magellanic Cloud (LMC), and Small Magellanic
Cloud (SMC) showcase  
how rapidly star-forming gas driven by
tidal interactions gets out of dwarf galaxies, eventually being accreted into larger
systems, in this case the MW.
They also show how the physical conditions of the 
interstellar medium (ISM) vary with heavy element abundance, gravitational 
potential depth, and radiation field.  The MW is in many ways an archetypal spiral galaxy, and so is an
important point of comparison for extragalactic studies. In addition,
there is some evidence from its star formation rate and global color
that the MW is undergoing one of the most important transitions
in galaxy evolution, leaving its period of rapid star formation and
entering the so-called ``green valley", and that it will have all but
extinguished its star formation in less than 5 Gyr \citep{mutch11}.
Studying the cold gas in the MW will elucidate how star
formation is quenched during this pivotal era.  Thus, it is of
fundamental importance to characterize the gas flow into, within, and
from the Galaxy.

The $\lambda$21-cm line is an excellent tracer of the neutral
interstellar medium (ISM) in galaxies. Atomic hydrogen (\HI) is found in
a variety of environments, from dense clouds to the diffuse galactic
halo and shows structure with size scales from kilo parsecs to a few
tens of AU.  Galactic \HI\ spectral lines vary in
width from $\sim 1$ \kms\ to $\sim 60$ \kms, tracing gas with
temperatures from $20\rm\,K$ to $8000\rm\,K$.  The SKA will combine surface
brightness sensitivity with the angular
resolution provided by long baselines to deliver a MW gas
survey machine. Combining SKA \HI\ data with
single-dish or auto-correlation data for the so-called
``zero-spacing'' will probe \HI\ in the MW and Magellanic
Clouds (MCs) over its full range of size scales and
temperatures.  

The combination of excellent \HI\ surface brightness sensitivity and
dense coverage of \HI\ absorption measurements will enable
measurements of
the multiphase gas flows into and out of the Milky Way disk and MCs.
In combination with tracers of molecular gas and dust, such as maps
made by Planck and
Herschel, these \HI\ data will reveal the transitions between atomic
and molecular material within interstellar clouds that are crucial
for star formation.
Only the SKA will be able to capture these clouds in the MCs for the
first time at sub-parsec resolution, providing important tests of the
effect of metallicity, dynamics and radiation field on molecular cloud formation
and evolution.  Over the next decade, GAIA and the Large Synoptic Survey
Telescope (LSST) will provide stellar distance measurements for constructing
4D (3D plus velocity) maps of nearby nearby interstellar clouds. Moreover,
the SKA will make it feasible to perform large-scale absorption
surveys of the OH molecule. It has recently been realized that this
molecule may be an ideal tracer of the so-called CO-free molecular
gas, thus allowing us to investigate in detail the transformation of
gas from atomic to molecular.

In this chapter we outline a number of areas where the SKA
will transform our understanding of how the MW and Magellanic
System work. This understanding will provide much of the physical
underpinning for how large spirals and dwarf irregular galaxies evolve.

\section{The Structure and Evolution of Gas in the Milky Way and
  Magellanic System}
\label{sec:MW}
\subsection{Understanding how gas accretes onto, moves within, and is lost from a galaxy}
Galaxies are not closed systems. The evolution of the MW is
significantly impacted by the two-way flow of gas and energy between
the Galactic disk, halo, and intergalactic medium. We have long known
that there is an extended halo of gas, both atomic and ionized, far
beyond the disk of the Galaxy
\citep[e.g.][]{lockman84,kalberla08,reynolds91,gaensler08}. In recent
years we have also come to realize that the atomic portion of the halo
is a highly structured and dynamic component of the Galaxy.  Despite
these advances we are far from understanding the origin of the gaseous
halo and its interaction with the disk of the Galaxy. There are two
dominant sources of structure in the halo: one is the outflow of gas
from the Galactic disk, and the second is the infall of gas from
extragalactic space \citep{putman12}.  The interplay of these sources
and their relative importance on the global evolution of the MW, and MW-like galaxies, is not known.

\begin{figure}
\centering
\includegraphics[width=12.5cm]{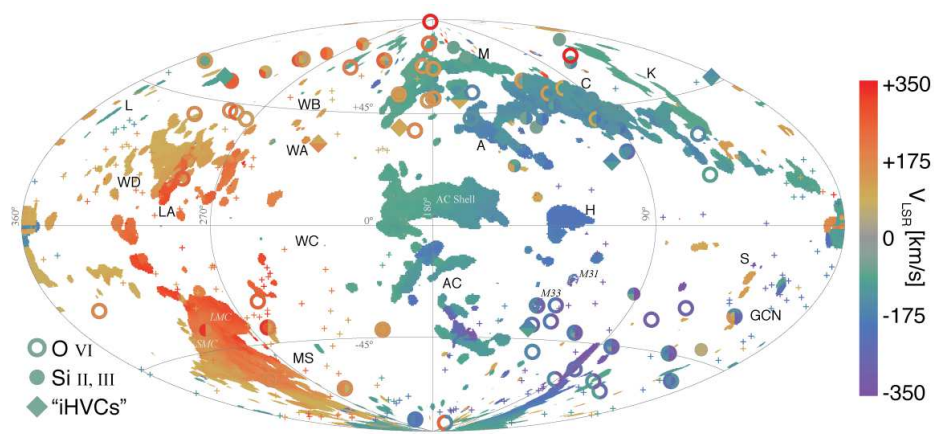}
\caption[]{Distribution of high velocity clouds (HVCs) detected in
  \HI\ and overlaid with
  measurements of low column density absorbers traced via ultraviolet
  metal lines (circles and diamonds).  The plus symbols represent
  compact HI clouds, the distribution of which will change
  dramatically with SKA.  The color scale
  indicates the local standard of rest velocity of the gas as shown in
  the bar at the right.  From Putman, Peek \& Joung (2012).
\label{fig:HVCs}}
\end{figure}

\subsubsection{Accretion}
\label{subsec:accretion}
Cosmological simulations predict that gas accretion onto galaxies is
ongoing into the present epoch \citep[e.g][]{joung12}.  Maintaining the star formation rate measured in galaxies across cosmic time requires significant amounts of gas infall
\citep[e.g.][]{hopkins08}.  The Milky Way is a clear example of this.
With $\sim 5 \times 10^9~{\rm M_{\odot}}$ of gas in the disk and a
current star formation rate of $1-3~{\rm M_{\odot}~year^{-1}}$, it is
clear that the Galaxy would exhaust its supply of star-forming gas in only a few
Gigayears \citep{putman12}. The problem of providing new gas to facilitate star formation is not
just limited to the current epoch.  Measurements of stellar ages in
disk stars show that they have formed continuously over
the past $\sim 12$ Gyrs \citep[e.g.][]{bensby14} and
chemical evolution models can only reproduce the abundances
observed in old stars by assuming continuous accretion of low
metallicity gas \citep[e.g.][]{chiappini01,schonrich09}.

An outstanding source of low-metallicity star formation material for disk
galaxies is gas accretion from the disruption of neighboring dwarf
galaxies.  The Magellanic System, created from the interaction between
the SMC, LMC and the MW, provides the
closest example of galaxy fueling.  While the Magellanic Leading Arm
is believed to be closely interacting with the MW disk, the
northern tip of the MS is furthest from the MW
and contains a wealth of small scale structure \citep{stanimirovic08}.
Surrounding the neutral MS is a significant pool of
ionized gas, indicated by UV absorption lines, which may account
for as much as three times the mass of the neutral gas \citep{fox14}.
By imaging both the neutral and the ionized gas we will be able to model
how the gas is being broken down, and ultimately how it is accreting
onto the MW. To image the neutral gas we will need the
combination of surface brightness sensitivity, spatial coverage and
angular resolution provided by the SKA.  Detailed \HI\ images to a low
column density limit will allow us to study the physical and thermal
structure of the Magellanic System throughout the halo and probe the
interaction between the Magellanic \HI\ features and the MW
disk and halo.  The \HI\ images can be compared with three-dimensional
maps of the warm, ionized gas made with large-area, high kinematic
resolution integral field units on 8-m optical telescopes to show the
flow of warm gas relative to the disk.  These studies will reveal the
dynamical and thermal instability processes that are essential for
feeding external material into galaxies. The SKA will also allow us to
trace cold gas throughout the Stream by resolving, spatially and
spectrally, cold clumps, and allowing us to search for absorption
through the Stream. Finally, these studies will help us to probe the
physical properties of the halo itself, using the Stream clouds as
test particles in the halo.
 
Another potential source of accreted gas is the reservoir of \HI\ in
the form of high velocity clouds (HVCs, see Figure \ref{fig:HVCs});
however, these clouds fall about an order of magnitude short of what
is needed \citep{putman12}.  Some of the missing mass is in the form
of ionized gas \citep[e.g.][]{lehner12, fox14} and another,
undetermined, small fraction may be in the form of small, dense clumps
of optically thick material.  While the ionized gas is best traced by
UV and optical absorption lines, gas at $T<8000$ K is well-traced by
\HI.  Our understanding of the link between the \HI\ gas detected in
emission and warm gas detected in ultraviolet absorption will change
dramatically with the SKA.  While high velocity \HI\ gas detected in
emission has been estimated to cover $\sim$35\% of the sky, high
velocity gas detected in absorption using metal lines is found to
cover on the order of 80\% of the sky.  This indicates there is a
large reservoir of gas in the Galaxy's halo with column densities of
N(\HI) $< 10^{18}$ cm$^{-2}$ that remains to be detected in \HI\ emission
\citep[e.g.][]{lehner12}.  This is consistent with what has been found
in other galaxy halos \citep[e.g.][]{werk14}.  Very deep integrations
of \HI\ and spatial stacking have detected some of this diffuse
(N(\HI) $\sim 2\times10^{18}~{\rm cm^{-2}}$) medium as HVC envelopes
with FWHM $\sim 60$ \kms\ at the interface with the hot halo gas
\citep{nigra12}. This diffuse low column density material could be
important for cloud lifetimes and overall accretion onto the Milky
Way.  To detect this gas we require a combination of extremely high
surface brightness sensitivity and ``zero-spacing'' observations for
\HI\ mapping of all angular scales.  Combining these maps with
sensitive \HI\ absorption observations, the SKA will be able to {\em
  detect} and {\em map} both the cold and warm gas reservoirs of Milky
Way star formation fuel.

We also know that cool ($T<1000$ K) gas exists in the HVC population.
In some cases, \HI\ cores have been found to be associated with
optical and UV absorption line systems when synthesis observations are
completed \citep{ben-bekhti09}.  \citet{kalberla06} find that 24\% of
HVC sight-lines show narrow-line width ($\Delta v \sim 7$ \kms)
components consistent with multiphase structure in the condensed halo.
Evidence from higher resolution observations suggests that beam
dilution hides some of this cool gas (Moss 2014)\nocite{moss14}.
Furthermore, data from FUSE show H$_2$ absorption in more than a dozen
high and intermediate velocity clouds \citep{richter01}.
Unfortunately, most of what we know about the multiphase \HI\ in the
MW halo is derived from single dish observations, which lack the
spatial resolution to clearly detect small, cold clumps of \HI.  We
know very little about the temperature, distribution or quantity of
cool \HI\ in the halo.  At present there are only two measurements of
\HI\ gas excitation, or {\em spin}, temperature in an HVC
\citep{wakker91b,matthews09}.  Absorption line measurements with the
SKA, with its resolution and sensitivity, will comprehensively
determine the amount of condensed gas in the halo enabling direct
comparisons with theoretical and numerical models (e.g. Wolfire et
al. 1995, Joung et al. 2012).  Furthermore, estimates of the mass of
\HI\ in HVCs rely on the assumption of optically thin gas when
calculating column densities \citep{putman02,moss13}, leading to
underestimates of the total HVC mass in the halo.  Using NVSS
continuum source counts \citep{condon98} we estimate that with the SKA
we will measure hundreds of absorption components through HVCs, giving
estimates of \HI\ spin temperatures and opacity in HVCs and estimates
of the fraction of gas ``missed'' by optical depth assumptions.

\subsubsection{Outflow}
A significant fraction of the gas in the Galactic halo may also be
attributed to the outflow of structures formed in the disk, but
extending into the halo.  Large-scale \HI\ ``chimneys''
\citep[e.g.][]{mcgriff06a,pidopryhora07} supply hot, metal-enriched
gas to the Galactic halo via the ``Galactic Fountain''
\citep{shapiro76,bregman80} and also may be a dominant source of
multi-phase gas structure for the lower halo.  Dense, cool
condensations 
from expelled gas \citep[e.g.][]{avillez00} and the tops of \HI\ chimneys that break at
$z$-heights of $\sim$1 kpc \citep[e.g.][]{maclow89} may populate the
halo. 
Deriving the destruction timescales and physical properties (temperature, density) 
of multiphase fountain gas will inform our estimates of the survival times and journey histories for cold halo cloud populations.
High
sensitivity SKA observations of \HI\ in emission and absorption,
together with observations of molecular and ionized gas tracers, will
reveal the multiphase structure of gas outflowing from the disk.
Comparison of multiphase gas properties with heating and cooling
models \citep[e.g.][]{wolfire95a} will determine the pressure and
thermal structure of the lower halo of the MW.

Galactic fountain gas may even play an important role in seeding the
halo to trigger gas accretion from so-called hot-mode accretion.
\citet{fraternali13} have suggested that high-metallicity expelled gas
mixes efficiently with hot halo gas and triggers the cooling of such
gas in the lower halo. This cooled gas is observed as absorption
features \citep[e.g.][]{lehner12} and can efficiently accrete onto the
disk.  Fraternali et al.\ estimated that supernova-driven fountain
cooling produces a net gas accretion onto the disk at a rate of a few
${\rm M_{\odot}~yr^{-1}}$ and speculate that this mechanism explains
how the hot mode of cosmological accretion feeds star formation in
galactic disks.  This intriguing suggestion can be tested by combining
SKA \HI\ observations with UV/optical absorption line measurements.  A
sensitive SKA survey of \HI\ emission in the lower halo of the Milky
Way will reveal the predicted low column density ($<10^{19}~{\rm
  cm^{-2}}$) tails of Galactic fountain material.  These images,
combined with \HI\ absorption will directly associate cool and warm
\HI, measure temperatures, and trace its origins to the disk.

\subsubsection{The SKA and Gas Tomography}
The development of the SKA is happening in tandem with an ongoing
explosion of stellar data. The gaseous properties can be linked to the
dust and distance measurements made using detections of stars in the
halo of the MW and LMC from future large
optical surveys, such as Pan-STARRS and LSST.  By measuring the
colors of stars with enormous photometric surveys we are able to
build 3D tomographic maps of the dust distribution in the Galaxy
\citep{green14,schlafly14}. In many circumstances, especially in
low-density environments, \HI\ and dust trace the same underlying
distribution of material. For example, by looking at absorption lines
toward stars, there have been some measurements of distances to high
velocity gas \citep[e.g.][]{thom08}, and from those estimates of
accretion rates onto the Galaxy \citep{putman12}. Thus, if we are able
to combine our kinetic information from a high sensitivity SKA
Galactic \HI\ survey with the tomographic information from stellar
surveys we will be able to produce truly 4D (3D plus velocity) kinetic
tomography maps of gas flow to and from the Milky Way, revealing how
feedback works in disk galaxies.  

Within our Galaxy's disk, structures are formed through the collapse
of diffuse material. By combining \HI\ kinematics with information on
the distance of gas, observations of \HI\ represent a direct
measurement of one of the most fundamental questions in all astronomy,
``where did this come from''?  To understand formation using observations of diffuse gas we look to the continuity
equation, which relates the rate of change of density ($\delta
\rho/\delta t$) to the spatial and kinematic state of the fluid.  The
position-position-velocity (or hyper-spectral) data provided by radio
surveys is clearly incomplete for this task --- we do not have access
to distance information. This leads to highly model-dependent
descriptions of the state of the Galaxy. As an example, maps of the
\HI\ surface density of the disk must rely on assumptions of a flat
rotation curve to extrapolate into the third spatial dimension
\citep[e.g.][]{levine06}. Even with these assumptions, we lose
information about flows of the gas beyond simple rotation, and thus
are blind to the more complex dynamics at play. The high angular
resolution \HI\ data cubes that we will obtain with the SKA can be used in
conjunction with deep photometric stellar surveys to finally enable 4D
kinetic tomography maps of gas flow within the MW. Combining
high angular resolution \HI\ data, which provides the information we
need to isolate individual \HI\ structures, with next-generation
photometric surveys of tens of billions of stars which give us precision
reddening information, will allow us to determine the distance to many
Galactic \HI\ structures  enabling reconstruction
of the detailed structure of the MW disk. With the full kinematic information,
vertical oscillations and corrugations of the MW disk will be related for the first time to
small-scale dynamical effects of spiral arms, giant molecular clouds, and/or sites of
stellar energy injection.

\subsection{Tracing the Life-cycle of Hydrogen from Diffuse to Molecular}

The current paradigm is that
gas in the MW is continually {\it flowing} between its various
thermal states.  The flow changes gas from its warm $T \sim 10^4$ K state,
through an intermediate unstable phase, to its cold atomic form, with
temperatures ranging from $\sim 20$ K to $500$ K.  
How exactly atomic gas transitions into  denser, colder clouds in
which the hydrogen is mostly in molecular form and in which stars form,
is still not understood.

The corollary of this hydrogen life-cycle is that the formation of stars ultimately destroys the
molecular clouds, and a large fraction of the gas is returned to
either ionized or atomic form.  Part of understanding how a galaxy
evolves is understanding how much hydrogen exists in its various
states and the flow rates between states and scales.  
The SKA is the only instrument capable of providing direct and statistically significant
measurements ($\sim 2 \times 10^5$ sources) of neutral gas states over the 
full temperature range from 20 to 10$^4$ K, while 
simultaneously probing diverse interstellar environments and spatial scales from AU to kiloparsecs.

\subsubsection{Distribution of Mass and Temperature in the Cool,
  Unstable and Warm Atomic Medium}

Interstellar gas is subject to strong radiative heating and cooling
processes which, within certain density or temperature ranges, may
cause a gas parcel to become depressurized when it is compressed
\citep{field65,field69,wolfire95a} leading to the so-called {\it
  thermal instability} and a runaway compression which lasts until
the parcel exits the unstable range.  This process tends to segregate
the atomic gas into a warm/diffuse phase 
called the warm neutral medium (WNM) and a cold/dense phase 
called the cold neutral medium (CNM). The
classical picture arising from this is that the medium should exist in
two or three main stable phases \citep{field69,mckee77} with very
little gas in the thermally unstable range. However, both
observational \citep[e.g.][]{dickey77,kalberla85,heiles03b} and numerical
\citep[e.g.][]{vazquez-semadeni00,gazol01,gazol05,audit05,hill12} studies
have suggested the presence of significant amounts of
unstable gas.  The existence of this unstable gas could be understood theoretically as a
consequence of several physical processes. Small-scale changes
in pressure caused by spiral arms or the distance from the Galactic
mid-plane can alter the temperature range where stable gas is observed.
Alternatively, the mixing action of turbulence, which causes the gas
to flow from one phase to another, can produce gas in the
traditionally unstable temperature range. However, the predicted
fraction of thermally unstable gas varies hugely across numerical models.
Direct comparison with observations over focused interstellar conditions
are highly required to constrain the above physical processes and timescales on which they operate.

The mass fraction in each of the hydrogen phase regimes, including the
unstable regime, is driven by the heating and cooling processes and
therefore is highly sensitive to parameters such as metallicity,
the interstellar radiation field and the strength of the turbulence
\citep{wolfire03}.  Surprisingly, even the fractions of the total
\HI\ mass in the CNM and WNM for most regions of the Galaxy
are poorly known and almost nothing is known about how they vary with
Galactic position.  Our knowledge of the unstable gas fraction is even
worse and the flow rates between stable and unstable phases are virtually
unknown. 

With the SKA we will achieve statistically meaningful measurements of
the distribution of mass and temperature in the CNM and WNM, as
well as the unstable phase. The SKA, with a deep survey of the
Galactic plane and lower halo, will provide those measurements through
detections of \HI\ absorption, probing spiral arms, inter-arm regions
and distance from the mid-plane.  For the first time we will be able
to measure gas spin temperature, 
distribution functions at all Galactocentric radii, as well
as 
in and above the plane, and close to some major Giant Molecular Clouds
(GMCs) with unprecedented spatial coverage of absorption sources. With this we
can 
test the theoretical predictions
on how the spin temperature varies with metallicity, interstellar
radiation field and pressure for the first time.

\subsubsection{Molecular cloud formation}
Recent numerical studies suggest that cold, dense clouds of atomic
hydrogen form by a phase transition from the WNM to the CNM
induced by large-scale compressions in the warm gas
\citep{ballesteros-paredes99, audit05, heitsch05,
  heitsch06,vazquez-semadeni06}. Such compressions may be a
consequence of the general turbulence in the WNM or by
large-scale instabilities that drive converging flows in this medium.
Cold clouds formed by this mechanism can eventually develop column
densities large enough for molecules to form, and ultimately form
stars.  Simulations of the formation of GMCs are rich and show that
GMC formation can happen via many different avenues.  One possibility
is that cold clouds develop gravitational instabilities, which lead to
contraction.  Thus, molecular gas and stars may be the result of
gravitational contraction starting in the atomic phase
\citep[e.g][]{vazquez-semadeni07,heitsch08a, heitsch08b, heitsch09,
  banerjee09} which, along the way, also produces complicated
filamentary structures \citep{gomez13}. As the gas flows from the WNM into the dense molecular structures, transitional regions that
are mostly cold atomic in their outermost parts, then CO-free
molecular, and finally CO-bearing molecular in their innermost parts are
produced \citep{smith14,heiner14}.  Alternatively, GMCs may develop
from an agglomeration of small cold clouds, collected together by the
sweeping action of spiral arms \citep{dobbs12} or through the
compressive action of superbubbles whose walls collect and compress
small \HI\ clouds \citep{ntormousi11,clark12}.

The detail apparent in the simulations is not yet matched by observational
data.  While undoubtedly most of the simulated physical processes are
involved in forming molecular clouds, we do not have the observational
handles to determine which processes dominate in different
environments.  Observing the flow pattern of cold atomic gas is
difficult because of the confusion in the atomic gas along lines of
sight inside our Galaxy. Dominated by turbulence and the overlay of
structures, the spectral structure of Galactic \HI\ line profiles is
usually wider than 7 \kms.  By comparison, typical molecular line widths,
although supersonic, are mostly on the order of $\sim$ 1 \kms. This
mismatch of spectral characteristics is the main reason why the atomic
to molecular transition in ISM has not been mapped systematically, in
spite of the existence of both large scale \HI\ maps and large scale
CO maps for more than three decades.  The observational challenge
lies in objectively associating flux in the \HI\ line with
a specific molecular structure, which is orders of magnitude smaller
both in spatial extent and turbulence magnitude.  
SKA \HI\ absorption measurements, combined with high surface
brightness sensitivity data at high Galactic latitude and sub-arcminute
resolution will generate a substantial catalogue of
isolated \HI\ clouds with simple Gaussian line profiles. These
clouds can be observed with matching angular ($\sim 20$'') and
velocity ($0.3$ \kms) resolutions in molecular tracers, such as OH, CO and
HCO$^+$. The matching of molecules and atoms will enable quantitative
analysis of important atomic to molecular transitions in many well-defined systems.

The SKA will finally give us the ability to observe the flow of cold
\HI\ onto molecular clouds and the mixture of phases within molecular
clouds.  One important tool in studies of cold gas flow is the
wide-spread effect of \HI\ Self-Absorption (HISA), which is observed
when cold foreground \HI\ absorbs background \HI\ emission at similar
velocities.  Many of these features are completely without molecular
counterparts \citep{gibson05a}.  Within molecular clouds, the
collision with H$_2$ produces a much colder and generally narrower
feature often called HI Narrow Self-Absorption (HINSA; Li \& Goldsmith
2003)\nocite{li03}.  HISA and HINSA allow us to {\em map} the
structure of cold gas, but because the temperature of the background
emission is unknown we cannot know the temperature of the cold gas.
By contrast, \HI\ absorption toward continuum sources gives us the
spin temperature of the gas, $T_s$, but not the spatial
structure. Using the SKA we will finally be able to use a combination
of HISA and \HI\ continuum absorption to understand the {\em spatial}
and {\em thermal} structure of cold \HI.  An SKA survey of the
Milky Way will enable sensitive absorption studies toward background quasars
giving opacity measurements with source density of $>10~{\rm deg}^2$.  This survey will provide
multiple absorption spectra across any given HISA feature, effectively
providing a temperature gauge to calibrate the HISA optical depth.
This powerful technique will provide accurate measurements of gas
temperatures and column densities around and within molecular clouds.
The SKA data will be combined with detailed studies of individual
molecular clouds from ALMA for a complete picture of the flow of cold
gas into and through molecular clouds.

The SKA will also enable essential complementary surveys of the hydroxyl
(OH) molecule, whose emission is very weak, despite its probable ubiquity and large observed column densities.  Recent evidence that a large fraction of molecular gas exists in the so-called "dark
molecular gas" (or CO-free) form \citep{grenier05,planck11,langer10}
amplifies the need to find other tracers
of cold H$_2$. In models of photo-dissociation regions (PDRs), OH
can form quickly in terms of extinction after $H_2$ becomes
self-shielded \citep{vandishoeck88}. A series of charge
exchange reactions facilitated by cosmic rays produces OH. Due to
its low excitation temperature, however, large scale maps of OH
emission has not been available.  A systematic and large scale OH
absorption survey, possible with the sensitivity afforded by the SKA, will quantify the
amount of this potentially most abundant diatomic molecule after H$_2$
and lead to a Milky Way-wide measurement of dark molecular gas. The
column densities and temperatures of OH will be the cornerstone to
build models of \HI-H$_2$ transition.

\subsubsection{The multi-scale SKA meets the multi-scale ISM} {\em
  Turbulence} is a concept often applied to random variations in the
density and velocity fields of the ISM.  The term
implies a stochastic process that transfers kinetic energy from larger
to smaller scales in a cascade similar to the well known Kolmogorov
process (Kolmogorov 1941, 1962, reviewed by Elmegreen and Scalo
2004,\nocite{elmegreen04} Falceta-Gon\c{c}alves et
al. 2014)\nocite{falceta-goncalves14}, although only a few studies
exist about the nature of the cascade in the strongly compressible
case (e.g., Kritsuk et al. 2007)\nocite{kritsuk07}.  Interstellar
turbulence affects many different kinds of observations, from spectral
line cubes to pulsar scintillation.  It is present in all phases of
the ISM, yet its exact energy sources and sinks are still not identified.  
Interstellar turbulence can be measured quantitatively either through the
structure function or power spectrum
\citep{rickett77,dickman85,lazarian00}.  The 21-cm line can be used in
emission or absorption to trace turbulence in both the WNM and CNM over a broad range of
scales.  The SKA will connect the \HI\ with turbulence in the warm
ionized medium (WIM) on large-scales as traced by pulsars
\citep{armstrong95} and polarization observations of the magneto-ionic
medium \citep{gaensler11}.  On small scales the SKA will connect the
CNM with turbulence in molecular clouds, where fluctuations in the
density and velocity fields are generally thought to shape the
processes of gravitational collapse and star formation (Larson 1981,
Qian et al. 2012; see also the review by McKee and Ostriker
2007),\nocite{larson81,qian12,mckee07} although recently it has been
proposed that, instead, these fluctuations are the result of global
gravitational contraction in the clouds rather than the other way
around \citep{ballesteros-paredes11}.

Spectral line cubes of the 21-cm \HI\ emission brightness over a large
area with a small beam are used to characterize the turbulence
spectrum over a range of scales, typically from $\sim$100 pc to
$\sim$0.1 pc \citep{chepurnov10,pingel13}.  The SKA will provide the
high sensitivity and large numbers of independent emission spectra needed to
compute the spatial power spectrum in the Galactic ISM down to AU
scales and with precision and statistical confidence better than in
previous studies \citep{crovisier83,green93,dickey01}.  In the
Magellanic System, the sensitivity of a deep SKA survey will give a
large improvement over the best existing data
\citep{stanimirovic01,elmegreen01,muller03}.  In nearby Galactic \HI\
clouds, moments of the spectral line cubes give the velocity and
density structure functions that can connect the turbulence spectrum
in the \HI\ to the very well known spectrum measured for the WIM
\citep{armstrong95}.  It will be particularly important to test
whether the neutral medium has structure similar to the extreme
scattering events (ESE) in the ionized gas \citep{fiedler87,walker98},
and to match structures in the \HI\ with scattering screens deduced
from pulsar scintillation \citep{bhat02}.  Additional statistical
measures such as those proposed by \citet{burkhart10} will probe
deeper into the fluctuations of the ISM.

In the CNM, variations in the optical depth and line centre velocities
among \HI\ absorption spectra toward nearby background sources will provide
a way of tracing turbulence independently of the \HI\ emission that is
dominated by WNM.  The SKA will provide orders of magnitude
more absorption spectra toward compact
continuum sources, making it possible to use these data to map the
variations in absorption on a wide range of scales.  This is
particularly important to connect with the ``tiny scale structure''
\citep{heiles97,deshpande00,braun05} that is seen in similar
variations of the absorption, particularly with VLBI (Roy et
al. 2012)\nocite{roy12}.  It will also be critical to link the turbulence in the CNM
to that seen in molecular clouds using CO and other molecular line
tracers \citep{falgarone09,krco08,li12} using ALMA.  Whether or not the turbulence
in molecular clouds is ``frozen-in'' with the gas when it makes the
transition from CNM atomic to molecular is an important question for
understanding star formation.  The multi-scale physics of the mass and
energy cascade from large galactic scales down to the small scales on
which molecular clouds shape the ISM is a challenge
for our understanding of galaxy evolution.  The SKA, with
its inherent spatial dynamic range, is ideally designed to probe the
multi-scale physics of the ISM.

\subsection{Taking ISM evolution to The Magellanic Clouds}
The MCs, including the SMC and LMC, offer a nearby example of a low-metallicity
($Z \sim 0.2~\rm Z_{\odot}$ and 0.5, respectively; Dufour 1975,
Olszewski 1996) environment with interstellar conditions that sharply
contrast with what we find in the MW.  Given their close
proximity ($\sim50$--$60\rm\,kpc$ Westerlund
1997)\nocite{westerlund97}, these are the only external galaxies where
we can study the atomic and molecular content-- from dense
star-forming regions to diffuse accretion streams-- at high (pc-scale)
spatial resolution.  In addition, the intense interstellar UV radiation
field of the SMC and the LMC, $4-10$ times higher than that in the
Solar neighborhood (Azzopardi et al. 1988), implies that heating and
cooling rates, dust-to-gas ratios, and chemical abundances in the MCs
represent the closest local examples of less-evolved systems common at
high redshift.  

\subsubsection{Interstellar environment and the fraction
of cold gas}
The census of cold gas and its conversion into stars over cosmic
time is one of several key parameters associated with galaxy
evolution; however little is known about the cold atomic gas even in very
nearby galaxies.  As the ``demography'' of cold gas and the phase mix
are largely driven by the heating and cooling processes -- whose rates
vary with metallicity, dust-to-gas ratio, and the strength of the
interstellar radiation field -- significant variations of the CNM/WNM
properties and abundances are expected from a theoretical point of
view \citep{wolfire03}.  The reality is that, in our home
neighborhood only 29 \HI\ absorption measurements exist for the cold
gas in the SMC \citep{dickey94,dickey00}, the nearest metal-poor
galaxy.  Similarly, only a handful of \HI\ absorption spectra exists
for the LMC (Marx et al.\ 1997), M31 \citep{dickey93}, and M33
\citep{dickey93,braun97}.  The only recent attempt to study properties
of cold gas in a lower-metallicity environment offered by the
outer radii of the Milky Way resulted in a highly puzzling result.
\citet{dickey09} suggest that the spin temperature of the CNM, and the
CNM fraction, stay constant with Galactocentric radius.  At a
Galactocentric radius of 25 kpc, where the supernova rate, metallicity
and interstellar radiation field are significantly lower than in the
inner Galaxy, the temperature of cold gas, contrary to all theoretical
predictions, is not different from what is found close to the Galactic
centre \citep{strasser07}.  The SKA will allow us to extend this
test to lower metallicities within the LMC and SMC.

The SKA will revolutionize our understanding of the cold neutral gas
in nearby galaxies, starting with the MCs.  With 25
background radio continuum sources ($S>20$ mJy) per square degree \citep{condon98},
$\sim2000$ \HI\ absorption spectra with $\sigma_{\tau}<10^{-2}$ will
be obtained towards the MCs.  This will allow us to measure the properties
of the CNM in the Magellanic Clouds and the mixture of warm and cool atomic gas,
as well as their respective spatial distributions.  A comparison of
the CNM/WNM statistics between the Milky Way and the MCs will show the
variation of the heating and cooling rates with metallicity and how
these processes affect the star formation rate.

\subsubsection{Formation of H$_2$ in galaxies}
Observational studies of galaxies
\citep{kennicutt98,bigiel08,schruba11} show that the surface density
of the star formation rate scales linearly with the surface density of
molecular gas.  This suggests that stars form in molecular clouds with
a relatively similar efficiency and therefore the ability to form
H$_2$ controls the evolution of individual galaxies.  Another
interesting observational result that came to light over the past few
years is that the surface density of atomic gas ($\Sigma_{HI}$) on
kpc-scales rarely exceeds $\sim10$ M$_{\odot}$ pc$^{-2}$
\citep[e.g][]{bigiel08}.  This saturation of $\Sigma_{HI}$ provides
important pointers for the physical conditions required to form
molecular gas out of the atomic medium.  For example, Krumholz et al.\
(2009)\nocite{krumholz09} showed that in the case of equilibrium H$_2$
formation, a certain amount of \HI\ surface density is required to
shield H$_{2}$ against photodissociation, and this \HI\ shielding
surface density depends primarily on metallicity.

\citet{lee12} found a relatively uniform \HI\ surface density of
6--8 M$_{\odot}$ pc$^{-2}$ for several dark and star-forming regions
in the Perseus molecular cloud in the MW and showed that $\Sigma_{HI}$ and the
H$_2$ fraction agree well with the equilibrium model predictions.
However, the key test for understanding H$_2$ formation across galaxies
requires observational probes of interstellar environments different
from those we find in the MW.  Extragalactic observations in the
pre-SKA era cannot resolve individual molecular clouds.  Only SKA
observations of the MCs will provide crucial tests for the \HI\ surface
density and H$_2$ fraction in low metallicity and high-interstellar
radiation field environments
on pc-scales required to test both equilibrium and non-equilibrium
\citep[e.g.][]{glover07} models for H$_2$ formation and molecular cloud
evolution.

\subsubsection{The warm neutral medium and excitation of \HI}

The WNM is one of the least understood phases of the ISM.  To
constrain theoretical and numerical models of the ISM and its
evolution over cosmic time, temperature distributions over the {\it
  full} temperature range from $\sim20$ to $\sim10^4$ K are essential.
This requires extremely high-sensitivity observations due to the very
low optical depth of the WNM, $\tau < 10^{-3}$.  Only two direct
measurements of WNM spin temperature exist for the MW so far
\citep{carilli98,dwarakanath02}.  Using the upgraded Very Large Array,
\citet{murray14} detected statistically the presence of a widespread
WNM population with $T_s=7200$ K.  This study was possible due to very
deep integrations, detection and modeling of the CNM lines, and then
stacking of the spectral residuals. This study demonstrates, for the
first time, that the non-collisional excitation of \HI\ is significant
even at high Galactic latitudes.  As Ly$\alpha$ scattering is the most
likely candidate for additional excitation of \HI, the Murray et al.\
results show that the fraction of Ly$\alpha$ photons, and/or the
photon propagation through the ISM, are likely more complicated than
what is currently assumed.  For example, both a theoretical study by
\citet{liszt01} and recent numerical simulations by \citet{kim14} 
assume a uniform flux of Ly$\alpha$ photons throughout the ISM and
result is the expected $T_s \lesssim 4000$ K.  Clearly, this discrepancy
between observations and theory needs to be understood.

The high sensitivity enabled by the SKA will allow direct measurements
of WNM spin temperature and its spatial variations for the first time.
As extremely high sensitivity is needed, the MCs are perfect
candidates for obtaining deep \HI\ absorption spectra ($\sigma_
{\tau}=10^{-4}$ per 1 \kms\ velocity channels) to provide a
statistically significant sample of the WNM.  With thousands of \HI\
spectra we will also be able to stack selectively to probe different
interstellar environments (e.g.\ close to major star forming regions
vs galaxy outskirts).

Finally, while the \HI\ content of the MCs has been
extensively studied alone and in synergy with other multi-wavelength
observations \citep[e.g.][]{kim98,stanimirovic99,leroy07,meixner13}, the
SKA will provide necessary sensitivity and angular resolution to match
\HI\ images with the resolution of Spitzer, Herschel (18$''$ and 12$''$,
respectively), and ALMA images.

\subsection{The Milky Way as a Foreground for Cosmology}
To properly measure the magnitudes and colors of populations of
extragalactic objects, a correction for the extinction and reddening
caused by dust in the MW is crucial. There are classically two
ways of mapping this correction across the high Galactic latitude
sky. The first is to measure the far-infrared (IR) emission from dust
particles and make an assumption about the ratio of this emission to
reddening and extinction. This method has been implemented by
Schlegel et al.\ (1998, SFD)\nocite{schlegel98}. The second is to assume
dust and neutral gas are well mixed, and that neutral gas dominates
the column density, so that the \HI\ column
density can be used as a proxy for extinction \citep{burstein78}. It
has recently been shown that known biases in the SFD dust map affect
large-scale structure observations to the point of significantly
biasing estimation of key cosmological parameters \citep{huterer13},
such as the evolution of the equation of state of dark matter
($dw/dt$) or cosmic non-Gaussianity. Thus, we require higher
precision, less biased dust maps than are currently available to make
best use of planned cosmological studies of large-scale structure. It
has also recently been shown that better dust maps can be constructed
by combining far-IR-based methods and \HI-based methods
\citep{peek13}. Thus, a high-resolution \HI\ column density map of the
high-latitude sky is a crucial product of the SKA.  At present it is not
known whether the CNM and WNM phases of the ISM have significantly
different ratios of reddening to integrated \HI\ line intensity, but
it is likely, as differing gas volume densities modify grain depletion
\citep{wakker00}. If this is the case, it is crucial that a high
latitude survey not only have high angular resolution and
sensitivity, but also high spectral resolution to get independent
estimates of CNM (narrow) and WNM (broad) columns for each line of
sight.

\section{Goals for SKA1}
Significant advances in our understanding of the evolution of the \HI\ in
the MW and Magellanic System from kiloparsec to AU scales can
be achieved through two surveys, both with high spectral and spatial
resolution.  One survey will be a shallow, high spectral resolution,
all-sky survey of Galactic \HI, which could be carried out commensally
with an all-sky continuum survey covering 1420 MHz.  The second will
target the Galactic Plane and Magellanic System with high sensitivity,
spectral resolution and angular resolution.  The two surveys are
described below.  Critical to both surveys are the requirements for:
a) high velocity resolution on the order of $\sim 0.3$ \kms\ to
resolve narrow spectral lines of $\Delta v\sim 2~{\rm km~s^{-1}}$, b)
excellent surface brightness sensitivity for imaging of diffuse \HI\
emission, c) capacity to recover ``zero-spacing'' flux by
combining interferometric and single dish data and d) significant long
baseline coverage to enable accurate \HI\ absorption measurements.

\begin{figure}
\centering
\includegraphics[width=2.95in]{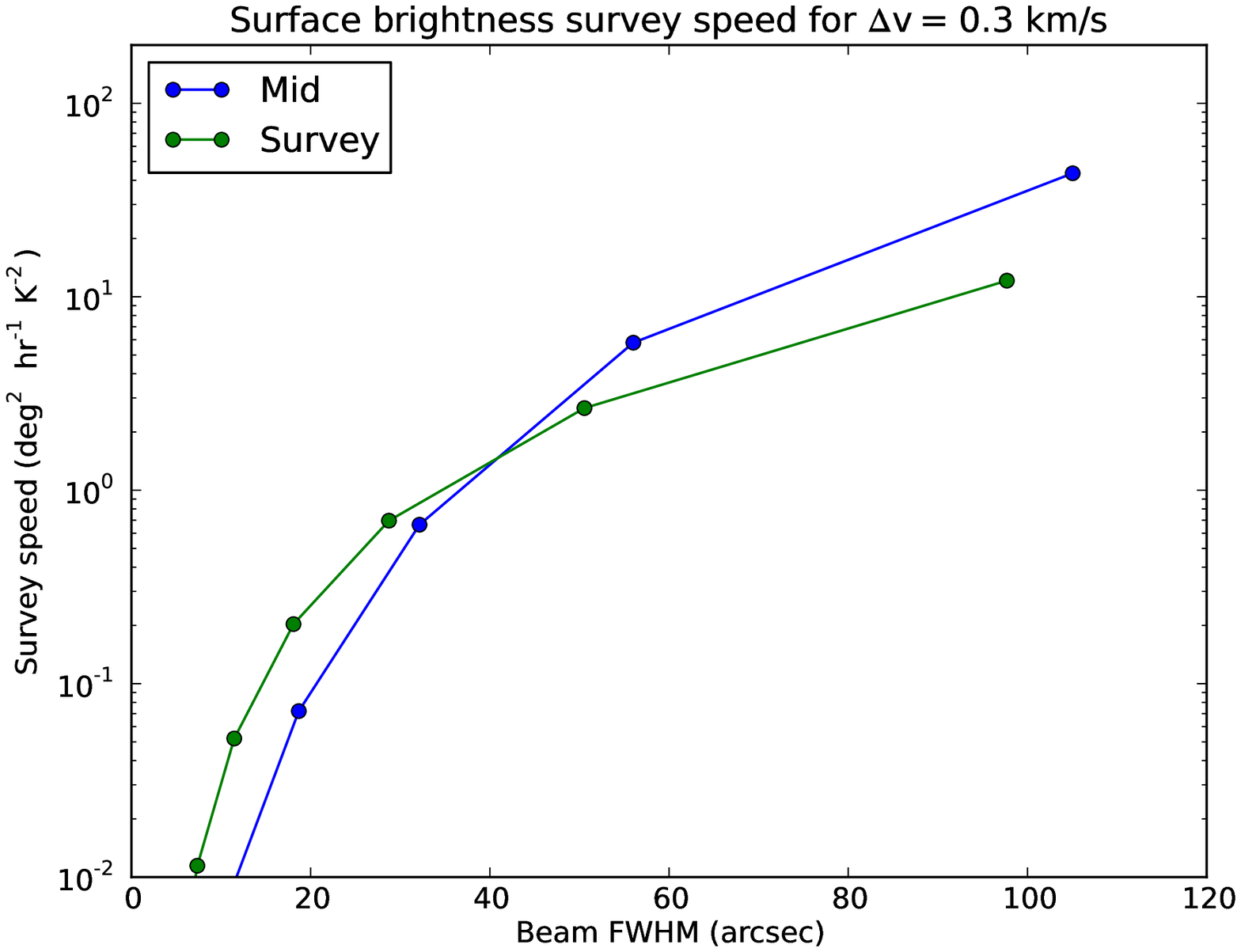}
\includegraphics[width=2.95in]{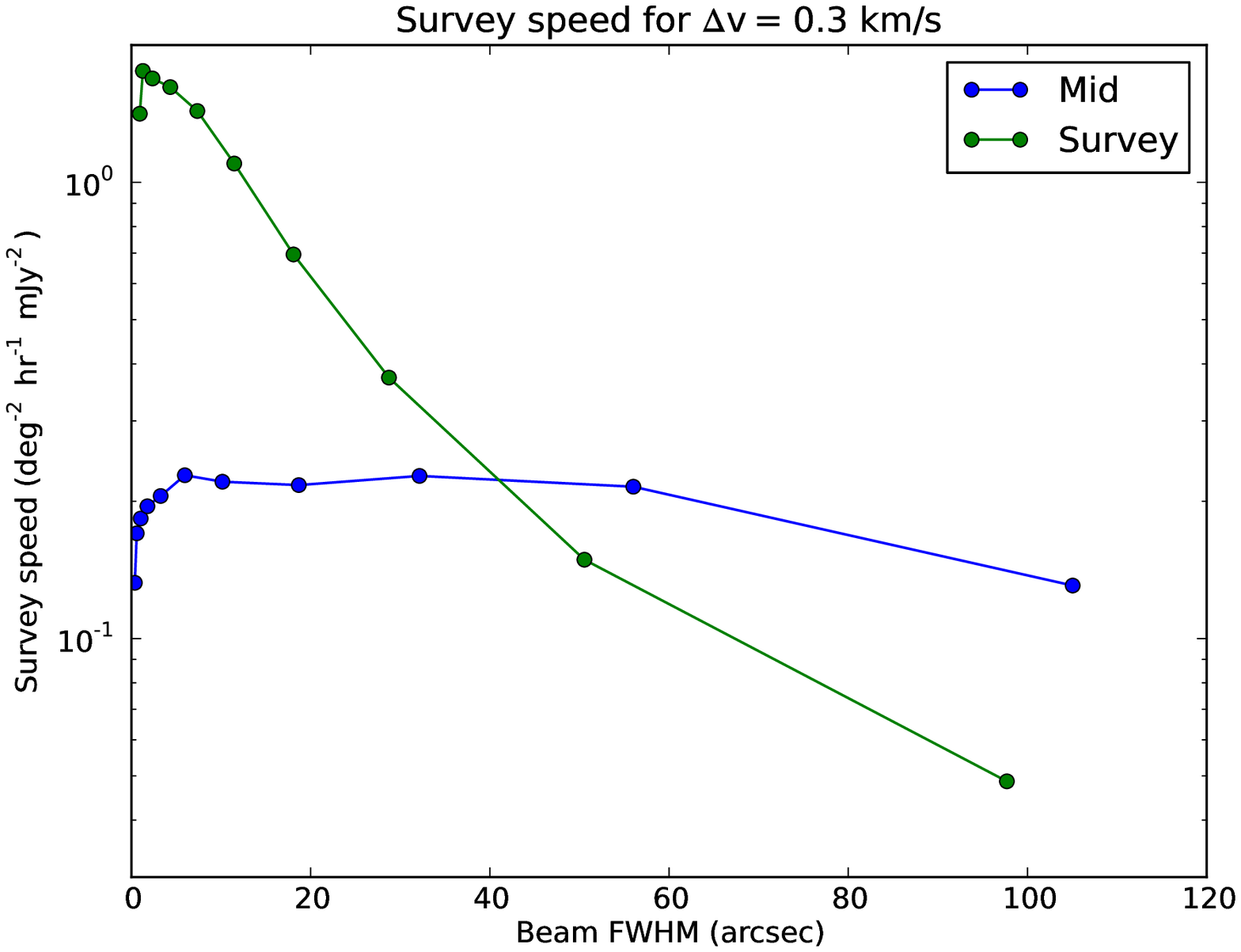}
\caption[]{Comparison of \HI\ line survey speeds for 0.3 \kms\
  channels on SKA1-MID and SKA1-SUR for {\em left}): surface
  brightness limited survey and {\em right}): flux limited survey.
  These are derived from simulations using the latest SKA1-MID and
  SKA1-SUR baseline distributions and assuming uniform weighting
  (Popping et al.\, 2014) and agree with the Braun (2014) SKA performance
  document.
\label{fig:speeds}}
\end{figure}

Brightness temperature limited survey speed as a function of beam size
for both SKA1-SUR and SKA1-MID is shown in Figure~\ref{fig:speeds}
(left) assuming a channel width of $0.3~{\rm km~s^{-1}}$.  Clearly for
heavily tapered data at angular resolutions of $>40$'' the survey
speed for SKA1-MID exceeds that of SKA1-SUR by up to a factor of a
few.  However, at the angular scales of interest for Galactic \HI\
($<30"$) the surface brightness limited survey speed for SKA1-SUR is
about a factor of 1.5 times that for SKA1-MID.  For measurements of
\HI\ absorption, the flux sensitivity at a spectral resolution of
$0.3$ \kms\ at the natural resolution of the array , $\sim 5$", is the
important quantity.  In Figure~\ref{fig:speeds} (right) we see that
for $\sim 5$" resolution the survey speed for SKA1-SUR is more than a factor of
10 faster than SKA1-MID.  To achieve the simultaneous goals of
Galactic \HI\ emission and absorption SKA1-SUR is clearly the optimum
instrument.   Assuming SKA1-SUR we estimate the expected brightness temperature
sensitivity, $\sigma_T$, with an angular resolution of 30'' and flux
density sensitivity, $\sigma_S$, at 5" per $0.3~{\rm km~s^{-1}}$
channel for three nominal integration times of 200 hr, 50 hr and 8 hr
(dwell time) as given in Table 1.

\begin{table}[ht!]
\centering
Table 1: Sensitivities for various dwell times for SKA1-SUR\\
\begin{tabular}{lccc }
\hline
\hline
Dwell time & $\sigma_T$ (30'')  & $\sigma_S$ (5'')  & Absorption
                                                      sources
                                                      (deg$^{-2}$)\\
\hline
200 hr         &       0.3 K   &    0.2 mJy     & 25\\
50 hr           &       0.7 K   &    0.5 mJy     & 14\\
8 hr        &       1.7 K   &    1.2 mJy     & 8 \\
\hline
\end{tabular}
\label{tab:surs}
\end{table}

\subsection{All-sky Survey of \HI\ Emission and Absorption with SKA1}
Using SKA1 we will  study
the interplay of
warm and cold atomic gas in the disk and halo by conducting a two-year
all-sky survey of \HI\ emission and absorption with SKA1-SUR.  We
estimate the observing times required to achieve necessary
sensitivities for the two components of this survey on SKA1-SUR.
Imaging of \HI\ emission will probe the diffuse circumgalactic MW and MS, 
enabling many of the scientific goals
outlined above.  A 2-year survey (8 hours per pointing) will give a brightness
temperature limit over the whole sky at 30\arcsec\ of about 1.7 K per
0.3 \kms\ channel, which is comparable to the most sensitive Galactic
plane interferometric surveys but covers the whole sky, including
the disk-halo interface and the Magellanic System, at a factor of
six better angular resolution.  This survey will probe
the structure of the circumgalactic gas, the MS and gas at the interface between the disk and halo.  Diffuse
emission requires significant numbers of short baselines for surface
brightness sensitivity and recovery of extended emission.
Furthermore, to understand the nature of the diffuse \HI, which fully
covers the sky we will need to include ``zero-spacing'' information,
as discussed in \S \ref{subsec:zero} below.

\begin{figure}[hb]
\centering
\includegraphics[width=4in]{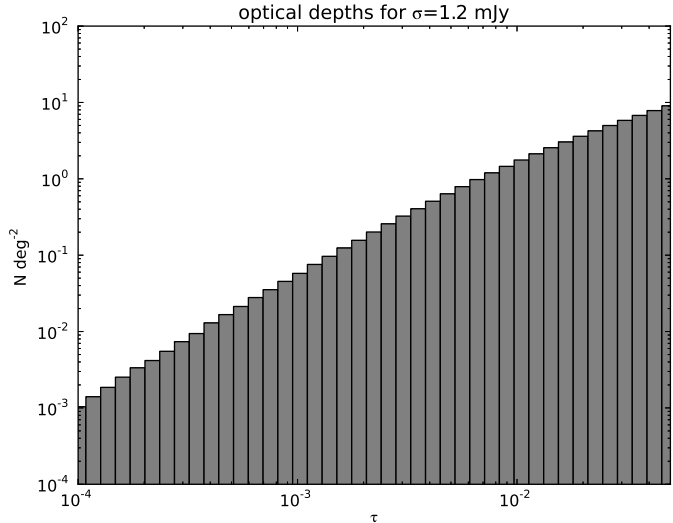}
\caption[]{Cumulative distribution of expected \HI\ optical depth measurements
  for an SKA1-SUR all-sky
  absorption survey.  The histogram shows the number of optical depths
  per square degree we expect to measure to a 
   given $\tau$ or smaller. These are based on detections in 0.3 km/s
  channels where $\sigma_S = 1.2$ mJy.  
\label{fig:tau}}
\end{figure}

The transformative aspect of the SKA1 all-sky survey will be the most
extensive study of \HI\ absorption in the Galaxy and Magellanic System
ever conducted, probing the physical properties of gas throughout the
Galactic halo.  A 2-year all-sky SKA1-SUR survey will give
$\sigma_S=1.2~{\rm mJy}$ in $0.3$ \kms\ spectral channels.  For \HI\
absorption, the flux density sensitivity converts to optical depth
sensitivity of $\sigma_{\tau} = \sigma_S / S_{bkg}$ for a given value
of the background source flux density, $S_{bkg}$.  Using
\citet{condon98} continuum source counts we estimate the number of
optical depth detections of $\sigma_{\tau}< 0.05$ per square degree in
Figure~\ref{fig:tau}.  The density of sources is also given  in column 4
of Table~\ref{tab:surs}.  We expect to obtain precision measurements
of $\sigma_{\tau}< 0.05$ towards $\sim 8~{\rm sources~ deg^{-2}}$ or a
total $\sim 2\times 10^5$ sources.

\begin{figure}
\centering
\includegraphics[width=13cm]{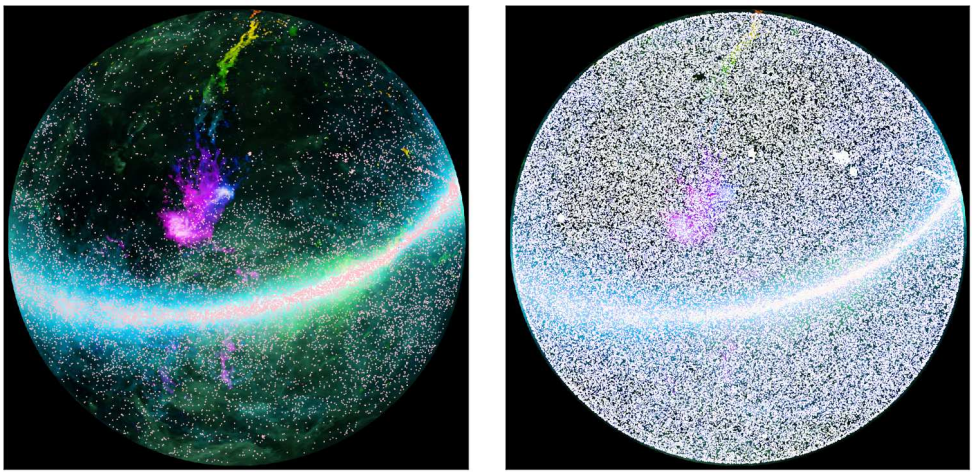}
\caption[]{Source density of \HI\ absorption measurements expected
  with SKA1 of the WNM (left) and the CNM (right).  Each
  dot is an anticipated absorption measurement.  The color image in the
  background is \HI\ emission, where color represents velocity from the Parkes
  Galactic All-Sky Survey \citep{mcgriff09}.
\label{fig:GASS_abs}}
\end{figure}

Figure \ref{fig:GASS_abs} shows the expected sky density of \HI\
absorption measurements based on $\sigma_{\tau}$ and weighted by the
\HI\ column density.  The left panel shows an estimate of the expected
number of WNM measurements with $T_s>4000$ K overlaid on an \HI\
total column density image and the right panel shows
an estimate of the number of CNM measurements with $T_s<500$ K.  The
accuracy of spin temperature measurements from \HI\ absorption is
limited by the error envelope imposed by emission fluctuations around
a continuum source.  It is therefore essential to have sufficient
surface brightness sensitivity on angular scales $<30$'' to measure
the fluctuations in emission.  Extrapolating spatial power spectra
currently measured in the Galactic plane and MCs
\citep{elmegreen01,dickey01} suggests that we will require
$\sigma_T<1.7$ K at 30'' to fully take advantage of the sensitivity of
the absorption measurements.

The average linewidth, even for cold \HI, will be 2.5 \kms, so some
gains can be achieved by averaging channels.  Averaging to $\sim1$
\kms\ channels will give $\sigma_{\tau}< 0.05$ towards more than 10
${\rm sources~deg^{-2}}$.  If $\tau = N_{HI}/ (1.8\times 10^{18}~{\rm
  cm^{-2}}\, T_s * \Delta v)$, we can detect $N_{HI} =
5.6\times10^{18}~{\rm cm^{-2}}$ in a 2.5 \kms\ line, assuming
$T_s=100$ K.  This will be an outstanding progression beyond the well
known Millennium survey \citep{heiles03b,heiles05}, which has been the
gold-standard for our knowledge of the distribution of temperatures in
the MW neutral ISM with its measurement of 202 cold neutral
components toward 79 high Galactic latitude sources.  

Furthermore, the all-sky SKA1 \HI\ absorption survey will measure
absorption in the WNM.  While the 21-SPONGE survey \citep{murray14}
will undoubtedly directly measure several WNM temperatures, the
numbers of measurements before the SKA1 are likely to be of the order
of tens.  The SKA1-SUR all-sky absorption survey will be able to
directly detect this WNM component towards hundreds of sources, as
shown in Figures \ref{fig:tau} and \ref{fig:GASS_abs}. Measurements of
\HI\ absorption will extend to HVCs, where there are currently only two
\citep{wakker91b,matthews09}.  Given the sky density of background sources and
the column density distribution of known HVCs (Moss et al.\
2013\nocite{moss13}), a blind all-sky survey covering all HVCs should
detect cold \HI\ components in absorption towards most known HVCs
($\delta < +30\arcdeg$) with narrow line components and measure their
spin temperatures.

For maximal efficiency of SKA1-SUR time this project should be
conducted commensally with an all-sky continuum or extra-galactic \HI\
survey, provided the correlator can be configured to provide 1.5 kHz
spectral channels over $\sim 5$ MHz around 1420 MHz.  This could be
achieved through either a ``zoom-mode'' or through narrow spectral
channels over the entire range PAF frequency range.  In this case, the
Galactic \HI\ absorption survey comes for free with the extragalactic
\HI\ survey.

\subsection{Targeted Galactic and Magellanic surveys}
To achieve the scientific goals of understanding the transition from
atomic to molecular gas described in \S \ref{sec:MW} SKA1 should be
used for a targeted survey of: ({\em i}) the MCs and
({\em ii}) the Galactic plane.  The required sensitivity for these two
regions is different, with the MCs requiring deeper
integrations over a comparatively small area.  Integrations of 200
hours on the MCs and 50-hours on the Galactic Plane will
exceed current surveys and those planned for the Australian SKA
Pathfinder (ASKAP) by about an
order of magnitude in sensitivity.  The flux limits for these surveys
are given in Table ~\ref{tab:surs}.

These surveys will deliver detailed \HI\ emission images capable of
measuring turbulence through the spatial power spectrum and probing
the structure of \HI\ flows around individual molecular clouds to AU
scales in the MW and pc scales in the MCs.  The data will be matched
to the resolution of data available from space missions, such as
Herschel, and molecular line measurements with ALMA.  Imaging of all
spatial scales, from 20\arcsec\ to many degrees, will be needed.  To
achieve this we will require short baselines as well as the ability to
combine single dish data, as discussed below.  With a spectral
resolution of less than 1 \kms\ our spatially resolved \HI\ images
will trace velocity gradients in gas flow.  As shown in
Table~\ref{tab:surs} the survey will deliver 14 \HI\ absorption
sources per square degree across the Galactic plane and up to 25 per
square degree in the MCs.  These measurements will allow comprehensive
measurements of the temperature distribution of gas across the Galaxy
and MCs as well as detailed calibration of the \HI\ self-absorption
observed in \HI\ emission images.  To successfully remove the effects
of \HI\ emission fluctuations from the absorption spectra we will
require $\sigma_T\sim 0.5$ K at 30''.

To achieve the needed sensitivity on the MCs ($100~{\rm
  deg^2}$) with 200 hours dwell time per point gives a total of $2
\times 10^4~{\rm deg^2~hr}$. The Galactic plane area ($|b|
< 2 \arcdeg$), is $1100~{\rm deg^{2}}$ so for 50 hours dwell time
per point the total is
$5.5 \times 10^4 {\rm deg^2~hr}$.  Dividing by the PAF FoV of SKA1-SUR
at 1.4 GHz, FOV= $18~{\rm deg^2}$ gives a total survey time of about
125 days.

\subsection{Zero-spacing information}
\label{subsec:zero}
Both the Galactic plane and all-sky surveys require the inclusion of
so-called ``zero-spacing'', to recover the total flux and structure
of the diffuse \HI\ emission that fills the sky.  The technical
aspects of how this might be achieved with the SKA have to be
considered carefully.  Typically ``zero-spacing'' data are provided from
single-dish surveys, which effectively sample baselines from 0m to the
size of the single dish.  Provided these data are matched in spectral
resolution and brightness sensitivity to the interferometric data and
the interferometer has sufficient baselines smaller than the size of
the single dish to allow for cross-calibration, the two datasets can
be combined to provide an image sensitive to all angular scales larger
than the resolution limit of the interferometer.  For SKA Phase 1 the
surface brightness sensitivity expected by even the deep Galactic
plane (0.3 K per 0.3 \kms\ channel) is at the level that can attained
by large all-sky single dish surveys similar to
GASS \citep{mcgriff09,kalberla10} and EBHIS \citep{winkel10}, which
both reach $\sim 50$ mK sensitivities at 1 \kms\ spectral resolution.
As SKA2 becomes available it may be necessary to use antennas of the
SKA itself, through its autocorrelations or other means, to obtain
short-spacings information at sufficient sensitivity. Although for SKA1 data can be obtained with existing single dish telescopes, it will be necessary to accommodate the data processing
requirements of adding short-spacings in all stages of the SKA.

\subsection{Phasing in of SKA1}
Galactic fields will benefit from the sensitivity of a fully scoped SKA1
1.  The Galactic demands pull the SKA design to its two extremes:
short baselines for surface brightness sensitivity and long baselines
for \HI\ absorption.  Given $\sim$50\% less collecting area, it would be
preferable to start with the shorter baselines as the survey speed for
surface brightness sensitivity scales as the inverse of the square of
the filling factor, or one over the longest baseline length to the
fourth power.  By contrast the survey speed for point source
sensitivity has no dependence on baseline length and goes simply as
the square of the collecting area.  The scientific objectives are such
that even for angular resolutions of $20$'' at 50\% of its baseline
collecting area SKA1 would be an advance over ASKAP. For a phased SKA1
working for some time without the longest baselines ($< 4000$m) is
satisfactory and could still provide an outstanding Galactic survey.
Furthermore, emission line surveys of the targeted Galactic plane and
MCs could be completed with worse angular resolution
than the natural SKA1-SUR array, and still achieve more than a factor
of five improvement in sensitivity at a factor of a few improvement in
angular resolution over ASKAP.

\section{All-sky H~{\sc i} Emission and Absorption with the SKA}
The ultimate goal for SKA is a complete survey of Galactic
\HI\ in emission and absorption over $3\pi$ steradians.  All-sky
Galactic \HI\ emission surveys have been limited to single-dish
surveys with angular resolutions of typically 10-60\arcmin\
\citep{winkel10,kalberla10,mcgriff09,kalberla05} and at best 3\arcmin
\citep[GALFA;][]{peek11}.  From these all-sky surveys comes the bulk
of our knowledge about the disk-halo interaction of the Milky Way, the
overall structure of the Galaxy and the distribution of the phases.
These provide essential foregrounds for interpreting observations of
everything from X-ray sources to cosmological backgrounds.  They have
been essential to recent estimates of the total gas quantities in warm
and cold \HI\ and dust from comparisons with Planck (e.g.\ Planck
collaboration 2013, Fukui et al.\ 2014)\nocite{planck13,fukui14}.  After very little
improvement in angular resolution over the past 30 years, SKA offers
us the first chance to conduct an all-sky MW survey at
interferometric resolution, improving the angular resolution by more
than an order of magnitude.  A fully equipped SKA with 10 times the
sensitivity of SKA1 will be almost unfathomably powerful, helping us
to reach sub-Kelvin sensitivity over the whole local \HI\ sky.
Finally we will have the sensitivity to map the low-column density
component of the high velocity sky in emission, while simultaneously
measuring the fluctuations in \HI\ emission necessary to interpret
\HI\ absorption spectra.  With these observations we will be able to
realize all of the goals outlined in \S \ref{sec:MW}.

Together with the all-sky \HI\ emission survey SKA will obtain
measurements of optical depths, $\tau <10^{-2}$ towards half a million
sources ($\sim 20~{\rm deg^{-2}}$; Figure 2), giving direct measurements of the
temperature and column density of the CNM
throughout the Galactic disk and halo.  Most excitingly, the SKA
all-sky HI absorption survey will be able to detect the warm WNM in absorption towards all > 700 mJy sources, giving $\sim
4000$ direct measurements of the temperature of the WNM.  With these
we will be able to determine the fractions of thermally stable and
unstable gas in differing galactic environments. Finally we will be
able to move beyond mapping to actually {\em measuring} the physical
properties of \HI\ throughout the MW and Magellanic System on
the angular scale that we currently observe \HI\ emission,
$15$\arcmin.


\section{Summary}
The SKA will revolutionize our knowledge of the evolution of the MW 
and the Magellanic System.  As we move first through SKA1
towards SKA2 we will be able to image the structure and measure the physical properties of \HI\ throughout
the Milky Way, its halo and the Magellanic System.  With these data we
expect to reveal the multi-scale physics that determines structure
formation in the MW, probe the transformation of atomic gas to
molecular clouds and determine the gas accretion processes
crucial to the long-term survival of the MW.  

The nature of structure formation is a question that drives many
fields of astrophysics, from the formation of galaxy clusters to the
formation of planets.  The processes driving structure formation are
multi-scale by nature with sources and sinks of energy and mass at all
scales.  The multi-scale nature of the ISM spans more than eight
orders of magnitude in spatial scale probing a tremendous range of
physical processes \citep{armstrong95}.  ISM studies with the multi-scale sensitivity of
the SKA will provide hereto inaccessible insight into the detailed
dynamical processes that govern the flux of mass and energy to and
from various ISM phases. These are crucial to understanding the hydrogen
cycle of galaxies, and in turn the evolution of structure in the
Universe.  

With the SKA we will have measurements of the thermal state, accretion
rates and origins of gas traveling into and out of the MW.
When combined with measurements of ionized gas we should finally be
able to produce a full census of material in the halo of the 
MW, and relate the rate and efficiency of accretion to the rate of
star formation within disk.

By studying \HI\ in three different galactic laboratories (MW,
Small and Large Magellanic Clouds) we will
reveal how \HI\ transforms into molecular clouds in regions with
different metallicities and UV radiation fields.  When compared with
theoretical models of molecular cloud formation this will have
important implications on the evolution of gas in a variety of cosmic
environments.

\bibliographystyle{apj}
\bibliography{references.bib}

\end{document}